\documentclass[prb, twocolumn, superscriptaddress, amsmath,amssymb]{revtex4}

\usepackage{graphicx}

\usepackage{dcolumn}

\usepackage{bm}

\begin{document}

\title{Optical Signatures of Momentum-Dependent Hybridization of the Local Moments and Conduction Electrons in Ce(Co,Ir,Rh)In$_{5}$ Kondo Lattices. }

\author{K.S.~Burch}

\email{kburch@lanl.gov}

\affiliation{Los Alamos National Laboratory, Los alamos, New Mexico 87545, USA}

\author{S.V.~Dordevic}

\affiliation{Department of Physics, The University of Akron, Akron, OH 44325}

\author{F.P.~Mena}

\affiliation{Material Science Center, University of Groningen, 9747 AG Groningen, The Netherlands}

\author{A.B. Kuzmenko}

\affiliation{D\'epartement de Physique de la Mati\`ere Condens\'ee, Universit\'e de Gen\`eve, quai Ernest-Ansermet CH - 1211 Gen\`eve, Switzerland}

\author{D. van der Marel}

\affiliation{D\'epartement de Physique de la Mati\`ere Condens\'ee, Universit\'e de Gen\`eve, quai Ernest-Ansermet CH - 1211 Gen\`eve, Switzerland}

\author{J. L. Sarrao}

\affiliation{Los Alamos National Laboratory, Los alamos, New Mexico 87545, USA}

\author{J. R. Jeffries}

\affiliation{Department of Physics and Institute for Pure and Applied Physical Sciences, University of California, San Diego, La Jolla, California 92093}

\author{E.D. Bauer}

\affiliation{Los Alamos National Laboratory, Los alamos, New Mexico 87545, USA}

\author{M. B. Maple}

\affiliation{Department of Physics and Institute for Pure and Applied Physical Sciences, University of California, San Diego, La Jolla, California 92093}

\author{D. N. Basov}

\affiliation{Department of Physics and Institute for Pure and Applied Physical Sciences, University of California, San Diego, La Jolla, California 92093}

\begin{abstract}
A new analysis of the optical properties of heavy fermion compounds is reported. We focus on the 1-1-5 series, where strong deviations of the spectra are seen from the predictions of the periodic Anderson model. Specifically we demonstrate that the differences between the experimental results and the theoretical predictions can be explained by accounting for  the momentum dependence of the hybridization between the local moments and the conducting carriers. Furthermore we find correlations between the hybridization strength on a particular band and some properties of the 1-1-5 compounds. These correlations suggest that the momentum dependence of the hybridization has to be taken into account, such that an understanding of the electronic properties of these heavy fermion compounds.
\end{abstract}

\maketitle

\section{Introduction}
\label{sec:intro}
The heavy fermion 1-1-5 series (CeTIn$_{5}$ where T=Co, Ir, or Rh) has attracted great interest lately since the discovery of  unconventional superconductivity in the vicinity of an antiferromangetic phase, reminiscent of the high temperature superconductors. These compounds reveal all canonical hallmarks of the heavy fermion state including large effective masses of the conducting carriers, the partially or completely screened Ce moment\cite{Hegger,Petrovic_CeCoIn5,Kim} and an optical gap-like feature that appears at low temperatures.\cite{Mena,Singley} The properties of heavy fermion compounds have generally been described by the mean field solution of the Periodic Anderson Model (PAM).\cite{Cox,Degiorgi} Within the PAM framework, mixing of local moments and conduction electrons leads to an enhancement of the mass of the charge carriers ($m^{*}$) and they in turn screen the local moments. This screening and mass enhancement is often referred to as the Kondo effect.\cite{Cox} Optical spectroscopy has proven to be an excellent probe of the key parameter of this model, namely the strength of the hybridization between local moments and conduction electrons ($V_{cf}$). Specifically optical studies have provided experimental access to the hybridization strength via direct observation of the hybridization gap ($\Delta$) in the dissipative part of the optical conductivity ($\sigma_{1}(\omega)$).\cite{Degiorgi,Dordevic,Hancock,Sievers,Dordevic_magfield,Dordevic_review, Gruner, Dressel} These studies have also extensively verified the prediction of the PAM on the scaling between the hybridization gap and the effective mass: $(\frac{\Delta}{T_{Coh}})^{2}\propto m^{*}$, where $T_{Coh}$ is the temperature below which coherent scattering between magnetic sites emerges.\cite{Dordevic} There have also been recent studies of numerous Yb and Ce compounds, that have demonstrated a scaling between the size $\Delta$ and $T_{K}$, the Kondo temperature.\cite{Hancock,Okamura} However, in almost all of the large number of studies of the optical properties of heavy fermion compounds, the focus has been on determining the size of $\Delta$ from $\sigma_{1}(\omega)$, and deviations of $\sigma_{1}(\omega)$ from the predicitions of the PAM  have generally been ignored. Here we report a new analysis of the optical conductivity data for the 1-1-5 series of compounds as well as additional measurements that provide insights into the role of the band structure and momentum (k) dependent hybridization in explaining deviations of the optical response from the line shape predicted by the PAM. Our results also suggest that k-dependent $V_{cf}$ governs a number of properties of the 1-1-5 family.

\begin{figure}
\includegraphics{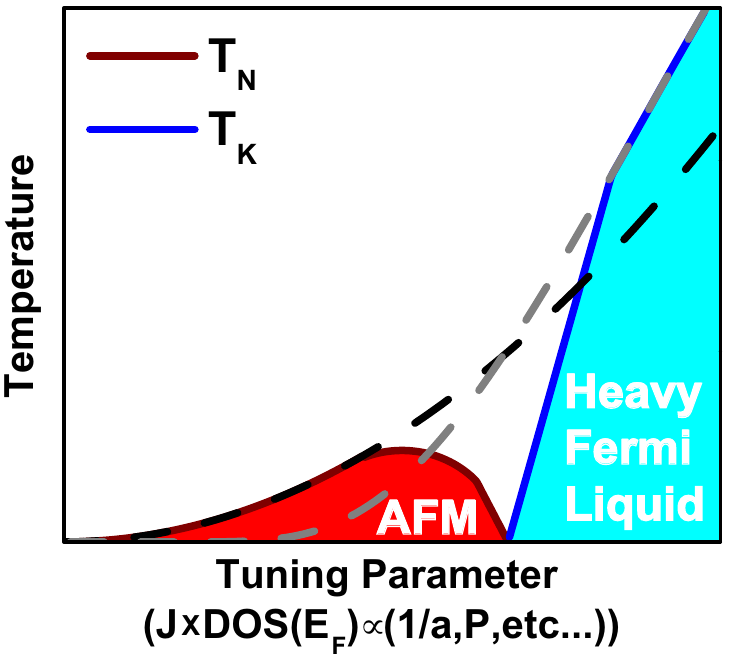}
\caption{\label{fig:doniach} The phase diagram originally proposed by Doniach for a one dimensional Kondo chain.\cite{Doniach_kondalattice} In this model the key parameter is the exchange ($J\propto\frac{V_{cf}^2}{W}$), where W is bandwidth of the conduction band, times the density of states at the Fermi energy ($DOS(E_{F})$). This parameter is generally believed to be tunable through the application of pressure (P), or by changing the lattice constant (a) via doping. At small values of $J\times DOS(E_{F})$  the RKKY interaction dominates and antiferromagnetic order  appears. As the tuning parameter is increased, the Kondo temperature ($T_{K}$) grows and eventually the antiferromagnetic order gives way to a heavy Fermi liquid at a quantum critical point.}
\end{figure}

Despite the success of the PAM, there are still a number of unresolved issues in the study of heavy fermions. One glaring problem is the difficulty with reconciling the coexistence of magnetically mediated superconductivity and the HF ground state. Indeed, within the PAM picture the Kondo effect should lead to a complete screening of the local moments, whose fluctuations are believed to trigger superconductivity in many unconventional superconductors.\cite{Mathur,Millis,Nakamura,Curro, Park, Sato, Jourdan, Steglich} Furthermore many HF superconductors, such as the 1-1-5 compounds, exhibit deviations from the mean-field predictions of the PAM.\cite{Kim} In particular, the Sommerfield coefficients ($\gamma = \frac{C}{T}$) often diverge at low temperatures, (ie: $\gamma(T\rightarrow0)\approx-ln(T)$ ), violating a key prediction of Fermi liquid theory ($\gamma \propto m^{*}$). Interestingly, it is generally believed that $m^{*}$ will diverge at a quantum critical point (QCP) separating magnetically ordered phases from a heavy fermion phase.\cite{Coleman} While a diverging $m^{*}$ may explain the enhancement of the Sommerfield coefficient in certain 1-1-5 compounds, it does not explain the non-Fermi liquid temperature dependence of the resistivity, and what connection this may have to the unconventional superconductivity found in these materials. Furhtermore it is a widely held belief that Doniach's picture of a quantum critical point separating magnetic and Kondo ground-states holds true in many HF compounds. In particular, he suggested that as $V_{cf}$ is enhanced the RKKY interaction that mediates a magnetic state gives way to the Kondo effect, such that a heavy Fermi liquid emerges when the magnetically ordered state is destroyed at a quantum critical point (see Fig \ref{fig:doniach}).\cite{Doniach_kondalattice} Therefore as the lattice constant is reduced, magnetic order should give way to a heavy Fermi liquid, that may exhibit superconductivity. This picture appears to explain the phase diagram of many heavy fermions, such as CeIn$_{3}$ an antiferromagnet with a Neel temperature ($T_{N}= 10~K$) that gives way to a heavy fermion superconductor at high pressures (ie: the lattice constant is reduced with pressure, increasing the strength of $V_{cf}$).\cite{Mathur}  However, despite their close relationship to CeIn$_{3}$, the 1-1-5 compounds do not appear to follow the Doniach phase diagram.

\begin{figure}
\includegraphics{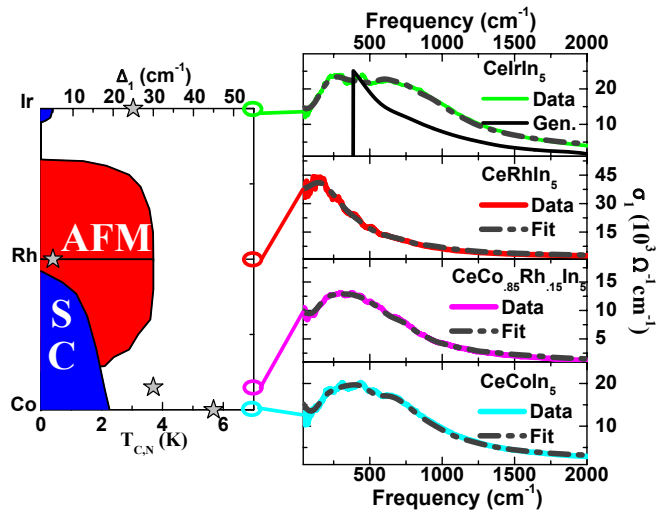}
\caption{\label{fig1} Left: The phase diagram of the 1-1-5 series as determined in ref. \onlinecite{Pagliuso02}, with brown stars indicating $\Delta _{1} $, the average value of the hybridization gap on a single band. The open circles indicate the region of the phase diagram probed by the $\sigma_{1}(\omega,T)$ plotted on the right. Each shows a broad MIR resonance, whose resonant energy becomes higher as one moves away from the Rh sample. The deviations of these results from the sharp resonance predicted by the PAM (black line) can be seen in the top right for CeIrIn$_{5}$. These differences are explained by the resulting $\sigma_{1}(\omega,T)$ using a distribution of gaps (P($\Delta$)) as shown by the dashed lines.}
\end{figure}

	In top left of Fig \ref{fig1} we present the phase diagram of the 1-1-5 series, details of which can be found in ref. \onlinecite{Pagliuso02}. We note that CeRhIn$_5$ is an antiferromagnet with a $T_{N}$ of 3.8 K, whereas the Ce moment is completely screened in CeIrIn$_5$ and CeCoIn$_{5}$, where heavy fermion superconductivity is found with a superconducting transition temperature (T$_{c}$) of 0.4K and 2.3K, respectively. Interestingly the in-plane lattice constant continuously decreases as one goes from Ir to Rh to Co, suggesting that the hybridization between Ce and In is roughly largest in CeCoIn$_{5}$, smallest in CeIrIn$_{5}$, and that $V_{cf}$ for CeRhIn$_{5}$ is between the two. Therefore from the theoretical phase diagram shown in Fig. \ref{fig:doniach}, one would expect antiferromagnetic order in CeIrIn$_{5}$ that gives way to superconductivity as one goes from Ir to Rh to Co, yet this is clearly not what is observed (see left side of Fig. \ref{fig1}). In addition, while the in-plane lattice constant is decreasing the c-axis lattice constant increases as one goes from Ir to Rh to Co, further complicating a comparison with the Doniach picture. Therefore to try to organize the phase diagram and address the underlying physical mechanisms leading to the complicated behavior of the 1-1-5 series we have employed infrared and optical spectroscopy, which, as mentioned above, provides direct access to the strength of hybridization. 

\section{Experimental Methods and Results}	
\label{sec:results}
	The high quality samples in this study were grown and characterized as described elsewhere.\cite{Hegger,Petrovic_CeCoIn5,Petrovic_CeIrIn5,Jeffries} Ellipsometry and infrared reflectometry were employed to determine the in-plane complex conductivity ($\hat{\sigma}(\omega)=\sigma_{1}(\omega)+\sigma_{2}(\omega)$) of all samples over a broad spectral range via the Kramers-Kronig relations as described in refs. \onlinecite{Burch} and \onlinecite{Kuzmenko}. The results for the CeCoIn$_5$, CeRhIn$_5$, CeIrIn$_5$ samples were originally obtained in Gronigen and presented elsewhere.\cite{Mena} To this data, at UCSD an additional sample (CeCo$_{0.85}$Rh$_{0.15}$In$_5$) was measured since it is believed to be close to the quantum critical point separating antiferromagnetic from superconducting order.\cite{Jeffries} Nonetheless the analysis of the data presented here is entirely new and has not been discussed previously. 
	 
	 We begin by exploring the evolution of $\sigma_{1}(\omega)$, the dissipative part of the optical conductivity, across the phase diagram of the 1-1-5 compounds.  The conductivity spectra at low temperature (7K) are displayed in the right panels of Fig. \ref{fig1}, noting that since the hybridization gap originates from strong correlations, it generally appears at $T<< \frac{\Delta}{k_B}$, where $k_B$ is Boltzman's constant.\cite{Burch,Degiorgi,Dordevic,Hancock} First we focus on the top right corner of Fig. \ref{fig1} where we show $\sigma_{1}(\omega)$ as predicted by the PAM (black line).\cite{Degiorgi,Dordevic,Hancock} The transition across the hybridization gap should have a sharp onset at $2\Delta$ followed by a rapid roll off. This is in stark contrast to the results for CeIrIn$_{5}$ (light green line in top right of Fig. 1) that contains significant deviations from the predictions of the PAM.  Specifically, the broad resonance centered at $500 ~cm^{-1}$ has a high-energy shoulder and no sharp onset. A similar response is seen in the other superconducting samples displayed in the bottom right, namely CeCoIn$_{5}$ and CeRh$_{0.15}$Co$_{0.85}$In$_{5}$. However focusing on the second panel on the right of Fig. \ref{fig1}, where we display the result for CeRhIn$_{5}$, we observe a clear mid infrared (MIR) feature that appears similar to the prediction of the PAM, yet no sharp onset is seen.
	 
	 Surprisingly, these results do not at first appear to produce any systematic trend with change in transition metal. Furthermore, it is tempting to simply attribute the broadening of the MIR hybridization gap feature to disorder. However judging from the relatively low value of the electrical resistivity, and long values of the electronic mean free path\cite{Petrovic_CeCoIn5}  one has to exclude the dominant role of disorder in broadening the gap structure in these compounds. As we will demonstrate below, the previous assignment of the MIR feature to excitations across a hybridization gap\cite{Mena,Singley} is correct. However, the exact form of the spectra and their variation within the series can only be quantitatively described by augmenting the PAM scenario with the momentum dependence of the hybridization strength. 
	 
\section{Analysis}
\subsection{Momentum Dependent Hybridization}
\label{sec:moment}
In order to appreciate the role of k-dependent V$_{cf}$  in governing the optical properties of the 1-1-5 compounds, it is imperative to analyze this parameter in the context of the electronic band structure. It is believed that there are four bands crossing the Fermi surface with different degrees of f-electron character in the 1-1-5 materials.\cite{Maehira} Furthermore, the band structure calculations indicate that this hybridization has significant k-dependence in each band,\cite{Maehira} which must result in a distribution of hybridization gap values in the system. Additionally, recent angle-resolved photoemission experiments on CeIrIn$_{5}$ have uncovered significant k-dependence in the degree of f-electron character at the Fermi energy.\cite{Fujimori} In order to account for the multiple bands crossing the Fermi surface and for the k-dependence of $V_{cf}$ we have calculated $\sigma_{1}(\omega)$ by taking the optical response to be a convolution of the result for a single hybridization gap\cite{Dordevic} ($\sigma_{1}^{PAM}(\omega,\Delta)$) with a spectrum of $\Delta$ values ($P(\Delta)$): 
  
\begin{equation}
\label{eq:sigma} \sigma_{1}^{dis}(\omega)=\int_{0}^{\omega_{c}}P(\Delta)\sigma_{1}^{PAM}(\omega,\Delta)d\Delta. 
\end{equation}
     
\begin{figure}
\includegraphics{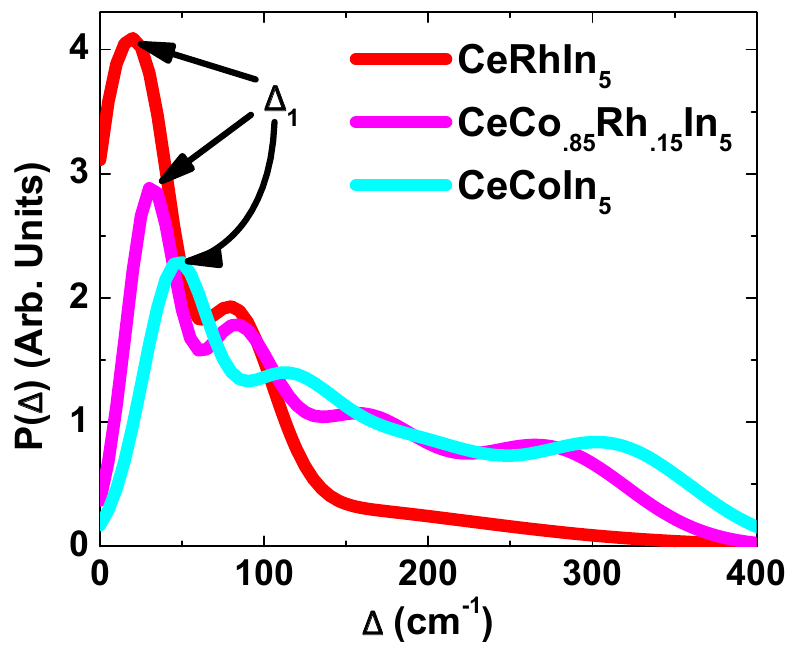}
\caption{\label{pdelta} The spectrum of hybridization gap values for three of the samples in this study. We note the significant increase in the average value of hybridization as one goes from Rh to Co.}
\end{figure}

The $P(\Delta)$ for three compounds and the resulting fits can be found in Figs. \ref{fig1} and \ref{pdelta}, respectively. Further details of the fitting procedure are outlined in the next subsection. One can think of this distribution as of a spectrum of the hybridization gaps on the Fermi surface, ie:  $P(E)\approx\int_{k_{f}}\delta(E-\Delta(k))d^{3}k$. As discussed in section \ref{sec:fit} we found that four Gaussians were needed to accurately represent $P(\Delta)$, which is reasonable since it is believed that four bands cross the Fermi surface.\cite{Maehira}

\subsection{Obtaining the Distribution of Gap Values}
\label{sec:fit} 
\begin{figure}
\includegraphics{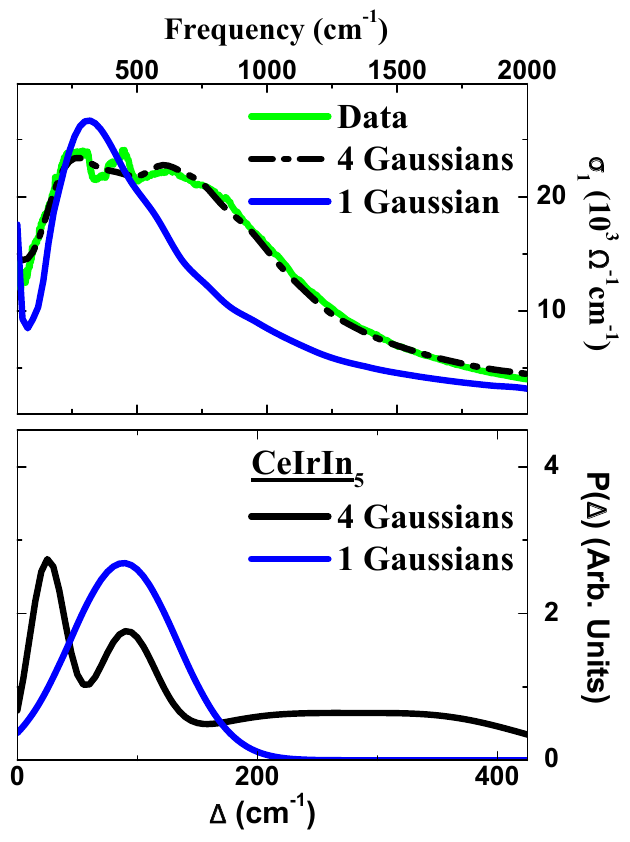}
\caption{\label{fig:fit} (Color Online) In the top panel the measured $\sigma_{1}(\omega)$ is displayed  in green along with the results of fitting the data with a one (blue) and four (black) Gaussians. The resulting distribution of hybridization gaps P($\Delta$) is displayed in the bottom panel when one (blue) and four (black) Gaussians are employed.}
\end{figure}

Using a spectrum of hybridization gap values to explain the optical properties of heavy fermions, is a new suggestion of this manuscript. However determining the proper P($\Delta$) for each sample was rather difficult as the formula for $\sigma_{1}(\omega)$ includes an integral (see Section \ref{sec:moment}), and therefore an analytic solution to determine P($\Delta$) is not readily apparent. Since P($\Delta$) results from the momentum dependence of the  V$_{cf}$, we used a Gaussian distribution to model P($\Delta$). The model was then adjusted such that the mean square error ($MSE=\sum_{i} (\sigma_{1}(\omega_{i})^{2}-\sigma^{dis}_{1}(\omega_{i})^{2}) $) was minimized as described in reference \onlinecite{Kuzmenko}. We initially attempted to fit the data using a single Gaussian in P($\Delta$), the result of such a fit for CeIrIn$_{5}$ is displayed along with the measured data in the top panel of Fig. \ref{fig:fit}. We have also included the response of the coherent carriers via the Drude formula: $\sigma_{1}^{Drude}(\omega)=\frac{\sigma_{1}(0) \Gamma}{\Gamma^{2}+\omega^{2}}$, where $\Gamma$ is the free carrier scattering rate, which was used as a fitting parameter and $\sigma_{1}(0)$ is the D.C. conductivity determined via resistivity measurements.

 From Fig \ref{fig:fit} one can see that the fit using a single Gaussian term does not reproduce the measured data well. In particular it does not result in a shoulder seen at approximately $700~cm^{-1}$ in CeIrIn$_{5}$. Furthermore a single Gaussian does not accurately reproduce the onset of the inter-band transition. However the use of a single Gaussian is still instructive. In particular despite the rather small energies that the Guassian covers in the bottom panel of Fig. \ref{fig:fit} ($0<\omega<200~cm^{-1}$), the result still effects $\sigma_{1}(\omega)$ over a large spectral range ($0<\omega<2000~cm^{-1}$). Therefore information about P($\Delta$) can be extracted at values of $\Delta$ much less then the measured range. We have ultimately determined that an accurate fit was only provided by four Gaussians, which, as discussed above, is in accord with the band structure calculations.\cite{Maehira} In particular the existence of the MIR shoulder and broad onset forced us to use four Gaussian distributions as is demonstrated in Fig. \ref{fig:fit}.

\subsection{$P(\Delta)$ results}
\label{sec:pdelta}
The results of our $P(\Delta)$ analysis uncover systematic trends across the phase diagram. This is seen in Fig. \ref{pdelta} where we display $P(\Delta)$ for three of the samples. One can clearly see that as Co is replaced with Rh, the weight of $P(\Delta)$ shifts to lower energies. Interestingly, this graph also suggests another significant change across the phase diagram, namely the size of the regions where the hybridization gap has collapsed (ie: $P(0) > 0$) grows significantly as Co is replaced with Rh. Since $\Delta$ is a measure of the hybridization strength (the PAM predicts: $\Delta(k)=\sqrt{E(k)^{2}+4V_{cf}(k)^{2}}$ , where E(k) is the dispersion for the conduction band), this result has significant implications for the magnetic moments in CeTIn$_{5}$. Noting that the hybridization gap is an experimental signature of the Kondo effect in the lattice, one can therefore infer that these nodal regions where  $\Delta=0$, indicate that some portion of the Fermi surface is no longer screening the local moments. This further suggests that the magnetic moment is only partially screened by the regions of the Fermi surface where $\Delta>0$ . Hence our findings of much larger nodal regions in CeRhIn$_{5}$ than in all other 1-1-5 compounds explains why the Ce moments in the Rh compound are only partially screened by the Kondo effect, whereas they appear to be completely screened in the superconducting compounds. Additionally, in Fig. \ref{pdelta} and the bottom panel of Fig. \ref{fig:fit} we also observe a distinct high-energy ($\approx 300-400~cm^{-1}$) peak in P($\Delta$) in the samples containing Co and Ir that explains the apparent shoulder seen at $\approx 600-700~cm^{-1}$ in $\sigma_{1}(\omega)$ for these compounds. Therefore we can explain both the origin of partially screened Ce moments and the deviations of  $\sigma_{1}(\omega)$  from the prediction of the PAM.  

\subsection{Underlying Order Parameter}
\label{sec:order}
In the phase diagram on the left of Fig. \ref{fig1}, we plot the center position of the lowest energy Gaussian of each compound ($\Delta_{1}$), which is a measure of the average value of the hybridization gap on a particular band. The value of this gap is also indicated in the distributions plotted in Fig. \ref{pdelta}. We note that the gap on this particular band is special since it contains nodes. Specifically, we find that the size of $\Delta_{1}$ is inversely proportional to the size of the nodal regions. This can be seen in Fig. \ref{pdelta}, by noting that as $\Delta_{1}$ moves toward zero, the size of P(0) grows. Therefore $\Delta_{1}$ will control the degree to which local moments can form and fluctuate. This suggests $\Delta_{1}$ is likely to be a fundamental parameter governing the physics of 1-1-5 materials, and may explain its apparent relationship to T$_{N}$ and T$_{c}$ seen in the phase diagram of Fig. \ref{fig1}. 

\begin{figure}
\includegraphics{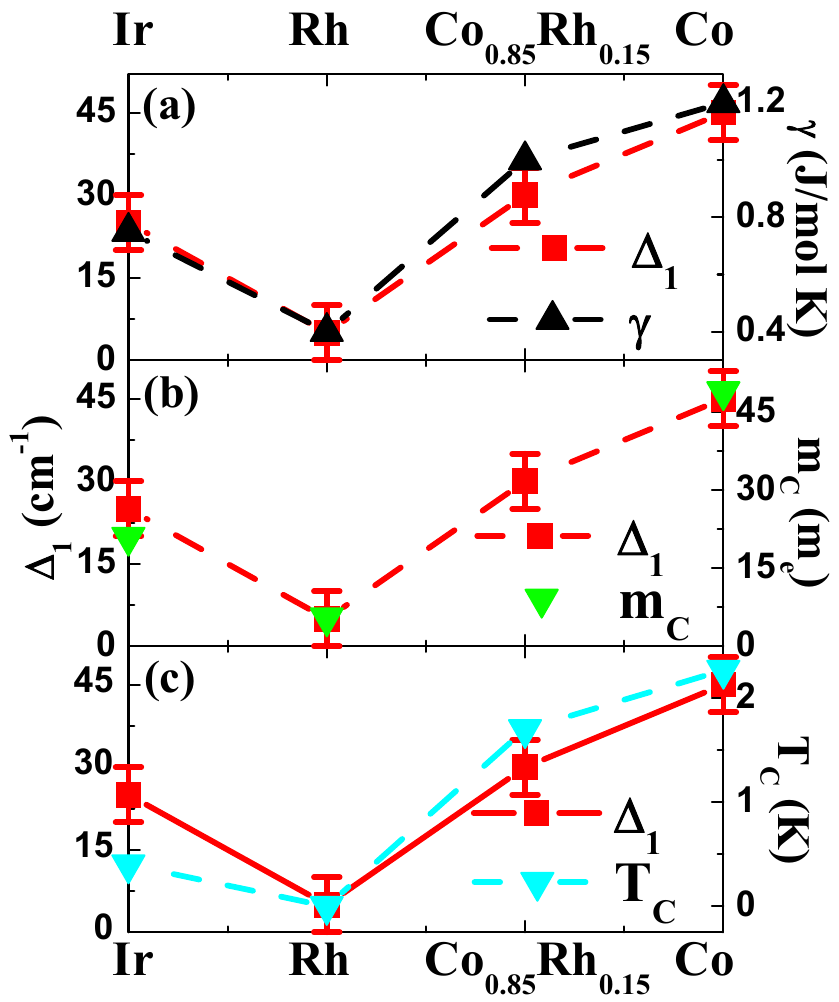}
\caption{\label{fig:correlations} The k-dependence of $\Delta$ for each sample is quantified by its distribution P($\Delta$). The average value for a single band is determined by the center position of the lowest energy Gaussian  ($\Delta_{1}$). This is shown along with (a) the Sommerfield coefficient  ($\gamma$), (b) the de-Haas van Alphen mass (m$_{c}$) for the 14$^{th}$ Hole band, and (c) the superconducting transition temperature (T$_{C}$).}
\end{figure}

If antiferromagnetic fluctuations are responsible for many of the properties of the 1-1-5 compounds, then it is important to explore what physical quantities $\Delta_{1}$ correlates with. In particular, Fig \ref{fig:correlations}(a) demonstrates that the normal-state Sommerfield coefficient ($\gamma$), as determined previously,\cite{Petrovic_CeIrIn5,Hegger,Petrovic_CeCoIn5, Jeffries} repeats the evolution of $\Delta_{1}$ across the phase diagram. One explanation for this correlation may be the smaller f-electron weight mixed into the Density of States (DOS) at the Fermi energy (E$_{F}$) implied by a smaller $\Delta_{1}$. To test this hypothesis, we have compared the de-Haas van Alphen mass ($m_{c}$) for the 14th hole band with $\Delta_{1}$ in Fig \ref{fig:correlations}(b), noting that this is the band that exhibits an enhanced mass near the antiferromagnetic quantum critical point.\cite{Haga,Settai,Shishido} In Fig \ref{fig:correlations}(c), we have plotted $T_{c}$ with $\Delta_{1}$ and also find that they are correlated. Surprisingly our attempts to correlate $\Delta_{2,3,4}$ with $\gamma$, $m_{c}$, and $T_{c}$ were unsuccessful. Furthermore we were unable to produce a reasonable agreement with the predictions of the PAM ($\Delta \propto \frac{1}{m_{c}}$).\cite{Coleman,Cox} Interestingly our preliminary tight-binding calculations\cite{Maehira} are consistent with the interpretation that $\Delta_{1}$ is from the 14th hole band. While this assignment is not unique, a more sophisticated many-body theory is required to confirm the connection between $\Delta_{1}$ and the 14th hole band.

\section{Discussion}
 In section \ref{sec:order} we have clearly shown that $\Delta_{1}$ appears to play a role in determining the properties of the CeTIn$_5$ compounds, yet the origin of this behavior remains unclear. In the Fermi-liquid picture $\gamma$  is related to the bulk DOS, as is $T_{c}$ in the Bardeen-Cooper-Schrieffer theory of supeconductivity. Therefore it is not clear why we find that $\gamma$ and $T_{c}$ only correlate with the hybridization gap of one band and are insensitive to the others. It is also unclear why we only find a correlation between $m_{c}$ and a corresponding $\Delta$ for only one of the bands, (namely $\Delta_{1}$ and not $\Delta_{2,3,4}$). This is especially strange since this band has the weakest hybridization strength (i.e. $\Delta_{1}<\Delta_{2,3,4}$), however this band displays the largest mass of all.\cite{Aoki,Haga,Settai,Shishido} Additionally, the rather large mass is believed to be related to the close proximity to a QCP, yet $m_{c}$ does not appear to diverge as $\Delta_{1}\rightarrow 0$. It is noteworthy that this happens to be the band that contains nodes in the hybridization gap. This suggests that the relationship between $\Delta_{1}$ and the properties of 1-1-5 compounds may be more complex. Nonetheless, the correlation between T$_{c}$ and the hybridization on one band is consistent with earlier findings that the superconductivity does not occur in all of the bands that comprise the Fermi surface.\cite{Settai,Tanatar}
 
In order to understand the importance of nodes in the hybridization gap, it necessary to ask how the properties of 1-1-5 compounds are affected by the k-dependence of the hybridization gap. Doniach has partially answered this question by showing that in the PAM, significant k dependence of $V_{cf}$ will still produce a Fermi liquid with either a Kondo or magnetic ground state.\cite{Doniach_kdependence} However, unlike the situation posed by Doniach, the hybridization gap in the 1-1-5 compounds contains nodes. We believe that these nodes are likely to produce complex behavior, remembering that  $\Delta_{1}\propto V_{cf}$. Therefore, nodes in the hybridization gap ensure the existence of some magnetic moment, allowing for the formation of antiferromagnetic fluctuations. Additionally, since the nodes only form on a fragment of the Fermi surface, these fluctuations can still strongly couple to the conduction electrons. The strong coupling of mobile charges to antiferromagnetic fluctuations is a likely pairing mechanism for superconductivity\cite{Curro,Mathur,Millis,Nakamura, Sato, Jourdan, Park,Steglich} and may also yield non-Fermi liquid behavior\cite{Varma}. This situation becomes even more intricate when the nodes develop mostly in one band crossing the Fermi surface. We therefore believe the existence of nodes in the 14th hole band explains why $\gamma$, $m_{c}$, and $T_{c}$ are correlated with $\Delta_{1}$, and may explain the non-Fermi liquid behavior displayed in these compounds. Furthermore, we note that these nodes and the apparent k dependence of $V_{cf}$ may explain the recent proposal of two fluid (Kondo impurity and Kondo lattice) behavior in the 1-1-5 series.\cite{Nakatsuji}

\section{Summary}
In this paper we have discussed deviations of the optical properties of the 1-1-5 series of compounds from the predictions of the Periodic Anderson Model. A new approach to understanding these differences was evaluated. Specifically, we have shown that the differences between the measured and predicted optical conductivity can be understood by taking into account the complicated band structure and momentum dependence of the hybridization that naturally leads to a distribution of hybridization gap values in the system. It is interesting to note that the approach proposed here should be applicable to other systems, yet the appreciable affect of k dependent $V_{cf}$ has been largely ignored in the vast studies of heavy fermions. Our results also provided useful insights into the origin of superconductivity and non-Fermi liquid behavior in this family of materials. Specifically,  $V_{cf}(k)$ for a particular band is clearly the tuning parameter governing the properties, and possibly defining the QCP of the 1-1-5 series that results in their non-Fermi liquid behavior. This is demonstrated by the correlation between $\gamma$, $m_{c}$, and $T_{c}$ and the average size of the hybridization gap on this one band. Moreover, it appears that the k-dependence of  $V_{cf}(k)$ allows for the formation of nodes in the hybridization gap that may select out this particular band for superconductivity. 	
	
\begin{acknowledgments}
We grateful for fruitful discussions with A.V. Chubukov, D.L. Cox, M. Di Ventra, Z. Fisk, J.N. Hancock, G. Kotliar, A.J. Millis, J.P. Paglione, D. Pines, J. Schmalian, and T. Timusk.
\end{acknowledgments}

\end{document}